\documentclass[review]{elsarticle}
\usepackage{mathtools}
\usepackage{amssymb}
\usepackage{bm}
\usepackage{subfigure}
\usepackage{multirow}
\usepackage{algorithm}
\usepackage{algpseudocode}

\usepackage{hyperref}

\journal{arXiv}

\begin{document}

\begin{frontmatter}

\title{A Simple Novel Global Optimization Algorithm and Its Performance on Some Benchmark Functions}


\author[mymainaddress]{YuanYuan Liu\corref{mycorrespondingauthor}}
\cortext[mycorrespondingauthor]{Corresponding author}
\ead{liuyuanyuan@mail.nwpu.edu.cn}

%

\begin{abstract}
This paper propose a new frame work for finding global minima which we call optimization by cut. In each iteration, it takes some samples from the feasible region and evaluates the objective function at these points. Based on the observations it cuts off from the feasible region a subregion that is unlikely to contain a global minimum. The procedure is then repeated with the feasible region replaced by the remaining region until the remaining region is ``small'' enough. If a global minimum is kept in the remaining region of each iteration, then it can be located with an arbitrary precision. The frame work is surprisingly efficient in view of its simple form and can be applied to black-box functions since neither special structure nor derivative information is required. The performance of the proposed frame work is evaluated on some benchmark functions and the results show that it can find a global minimum rather quickly. 
\end{abstract}

\begin{keyword}
global optimization, optimization by cut, benchmark functions 
\end{keyword}

\end{frontmatter}


\section{Introduction}\label{Sec1}
In this presented paper we focus on solving the problem
\begin{equation}\label{GOP}
\min_{x\in\Omega}f(x)
\end{equation}
of finding a global minimum of the objective function $f:\Omega\subset\mathbb{R}^D\to\mathbb{R}$ defined on the feasible region $\Omega$. The function $f$ may be highly nonlinear, non-convex, and possibly with multiple local minima. Moreover, we assume that neither special structure nor derivative information of $f$ is available, i.e., it is a block-box function. The function $f$ may even be discontinuous. For simplicity, however, we shall assume that the feasible region $\Omega$ is of simple form, say, a rectangular region $\left\{\bm{x}\in\mathbb{R}^D:\bm{x}_L\le\bm{x}\le\bm{x}_U\right\}$. But this assumption is not essential for our approach.

Generally, finding a local minimum is relatively simple and can be by achieved by various classical local algorithms, e.g., gradient descent or Newton's method \cite{boyd2004convex} if derivative is available or pattern search \cite{torczon1997convergence} if not. A straightforward way to generalize these classical methods to find a global minimum is to start multiple local searches from different initial points \cite{boender1982stochastic,byrd1990concurrent}. However, this method can easily be trapped into a flat region or a local minimum if the objective function has many local minima. Thus, one has to find methods that are global in its nature. Roughly speaking, the existing global optimization methods can be divided into two categories. The first category is the deterministic methods, including Lipschtiz optimization \cite{gaviano2008global,malherbe2017global}, branch and bound \cite{androulakis1995alphabb,adjiman1998global1,adjiman1998global2,epperly1997reduced}, Baysian optimization \cite{jones1998efficient,solak2003derivative,gardner2014bayesian,williams2006gaussian}, and so on. Lipschitz optimization assumes only the weakest (if there exists any effective algorithm) assumption one could expect on the objective function, i.e., it has a finite Lipschitz constant. But this approach can be inefficient if the Lipschitz constant is large. On the other hand, branch and bound make a full use of the special structure of the objective function and hence can be very efficient if tight lower bound can be constructed. However, this kind of special structure is not available in many applications, e.g., black-box functions. Bayesian optimization is tailored for functions that are very expensive to evaluate, it can find a global minimum with a few evaluations. Moreover, it requires neither special structure nor derivative information and hence can be applied to black-box functions. But in each iteration it has to maximize a so called acquisition function (see \cite{frazier2018tutorial} for details) to decide where to sample next. This acquisition function itself may be highly multimodal and very hard to optimize.  

The second category is the stochastic methods, including many heuristic algorithms such as genetic algorithms \cite{mitchell1998introduction,goldberg1988genetic,greenhalgh2000convergence}, evolution strategies \cite{schwefel1993evolution,beyer2002evolution,li2020evolution}, differential evolution \cite{price2006differential,das2010differential,deng2020improved}, and particle swarm optimization (PSO) \cite{bratton2007defining,kennedy1995particle,poli2007particle}. In these methods, an initial set of candidates (often called a population) is generated and iteratively updated through a mechanism that is inspired by biological evolution. Usually, each new generation is improved by a little bit with respect to its predecessor and consequently, the population may gradually evolve to a global optimum. Take the PSO for instance, a swarm of particles $\bm{x}_i=(x_i^1,\cdots,x_i^D)$ are originally initialized in a uniform random manner throughout the search space; velocity $\bm{v}_i=(v_i^1,\cdots,v_i^D)$ is also randomly initialized. The entire swarm is then updated at each time step by the following rules \cite{bratton2007defining}:
\begin{equation}\label{PSO_Update}
\begin{split}
v_i^d &= wv_i^d+c_1r_1(p_i^d-x_i^d)+c_2r_2(p_g^d-x_i^d)\\
x_i^d & = x_i^d+v_i^d
\end{split}
\end{equation}
where $\bm{p}_i=(p_i^1,\cdots,p_i^D)$ is the best position individually found by the $i$-th particle; $\bm{p}_g=(p_g^1,\cdots,p_g^D)$ is the best position found by the entire swarm; $r_1$ and $r_2$ are independent random numbers uniquely generated at every update for each individual dimension $d = 1$ to $D$; $w,c_1$, and $c_2$ are tunable parameters.
 
In this paper, we shall propose a new frame work, which we call optimization by cut, for solving the global optimization problem \eqref{GOP}. The philosophy behind is totally different from that of the existing methods. In the existing methods, a global minimum is obtained by successively constructing better solutions based on current observations, while in our method this is achieved by successively cutting off larger and larger subregions that are unlikely to contain a global minimum. It is inspired by the simple observation that if we take enough sample points from the feasible region, then a global minimum should be contained in a small region around the best point. At first glance, this simple idea seems to apply only to functions with a small number of local minima. However, it turns out that it also applies to functions with many local minima.

Compared to the PSO, the proposed approach has the following advantages:
\begin{enumerate}[(1)]
	\item In each iteration, the updating rule of the proposed approach is computationally simpler than that of the PSO \eqref{PSO_Update}.
	\item All the sample points generated by the proposed approach are contained in the feasible region while some of that generated by the PSO may not. 
	\item The proposed approach can be either in deterministic or stochastic form while the PSO can only be in stochastic form.
	\item The proposed approach usually converges faster.
\end{enumerate}
The disadvantage is that the proposed approach usually requires more evaluations of the objective function and is more likely to be trapped into a local minimum if the objective function has a high dimension and lots of local minima.

The remaining part of the paper is organized as follows. The proposed optimization by cut algorithm is presented in Section \ref{Sec2}. The benchmark functions for evaluating the performance of the proposed algorithm are collected in Sections \ref{Sec3}. Numerical results are given in Section \ref{Sec4}. Finally, the conclusions are presented in Section \ref{Sec5}.

\section{Optimization by Cut}\label{Sec2}

The idea behind the proposed optimization by cut algorithm is very simple and can be stated as follows. For a given function $f:\mathbb{R}^D\to\mathbb{R}$ with feasible region $\Omega\subset\mathbb{R}$, let $\Omega_0=\Omega$ and draw some samples $\bm{x}_1,\cdots,\bm{x}_N$ from $\Omega_0$. Based on the values of $f$ on these samples, we cut off from $\Omega$ a subregion that is unlikely to contain a global minimum of $f$ to obtain a ``smaller'' region $\Omega_1$. Now, the procedure can be repeated with $\Omega_0$ replaced by $\Omega_1$ until for some $n$ the remaining region $\Omega_n$ is ``small'' enough. If we can keep a global minimum of $f$ contained in the remaining subregions $\Omega_1,\cdots,\Omega_n$, then we can locate it with an arbitrary precision. This idea is summarized in Algorithm \ref{Algo1}.

\begin{algorithm}[t]
	\caption{Abstract Optimization by Cut}
	\label{Algo1}
	\hspace*{0.02in} {\bf Input:} $f:\mathbb{R}^D\to\mathbb{R}$ and $\Omega$
	\begin{algorithmic}[1]
		\State $\Omega_0\leftarrow\Omega$
		\For{$n=1,2,\cdots$}
		\State draw some samples from $\Omega_{n-1}$
		\State evaluate $f$ at the sampled points
		\State cut off a subregion from $\Omega$ based on current observations to obtain $\Omega_n$
		\State \textbf{quit} if $\Omega_n$ is ``small'' enough
		\EndFor
	\end{algorithmic}
    \hspace*{0.02in} {\bf Output:} $\bm{x}_{\min}$ be the optimal sample point and $f_{\min}=f(\bm{x}_{\min})$
\end{algorithm}

In order to implement the optimization by cut algorithm, however, we have to actually realize step 1 and 3 in Algorithm \ref{Algo1} and give a stop criterion explicitly. 

\subsection{Draw samples from $\Omega$}\label{Sec2.1}
There exist two kinds of sampling strategies, we can either draw samples from $\Omega$ deterministically or randomly. The simplest deterministic approach is to take the sample points as the vertices of a grid in $\Omega$. This approach is explicitly given in Algorithm \ref{Algo2}, where we have restricted our attention to the case of rectangular region for the sake of simplicity, i.e.,
\begin{equation*}
\Omega=\{\bm{x}\in\mathbb{R}^D|x^d_L\le x^d\le x^d_U,d=1,\cdots,D\},
\end{equation*}
where $x^d$ is the $d$-th component of $\bm{x}$ and $x_L^d$ and $x_U^d$ denote, respectively, the lower and upper bound of $x^d$. While this sample method is very simple, it can not be applied to high dimensional problems. Take the $30$-dimensional case for instance, even if we take $N=2$, the number of sample points would be $N^D=2^{30}\approx 10^9$ which is computationally intractable. More sophisticated approaches that can be applied to high dimensional problems are the so called space-filling designs such as Latin hypercubes (see \cite{mckay2000comparison} for details).

\begin{algorithm}[t]
	\caption{Draw samples deterministically}
	\label{Algo2}
	\hspace*{0.02in} {\bf Input:} $\bm{x}_L,\bm{x}_U$, and $N>1$
	\begin{algorithmic}[1]
		\State $\bm{y}_n\leftarrow\bm{x}_L+(n-1)/(N-1)(\bm{x}_U-\bm{x}_L),\quad n=1,\cdots,N$
		\ForAll{$n_d\in\{1,\cdots,N\},d=1,\cdots,D$}
		\State 
		$\quad$$\bm{x}_{n_1,\cdots,n_D}\leftarrow(y_{n_1}^1,\cdots,y_{n_D}^D)$
		\EndFor
	\end{algorithmic}
	\hspace*{0.02in} {\bf Output:} $\bm{x}_{n_1,\cdots,n_D},n_d\in\{1,\cdots,N\},d=1,\cdots,D$
\end{algorithm}

Alternatively, we can draw samples from $\Omega$ randomly. The simplest random approach is to make the sample points uniformly distributed on $\Omega$, i.e.,
\begin{equation*}
\bm{x}_n\sim\mathcal{U}(\Omega),\quad n=1,\cdots, N
\end{equation*}
where $\mathcal{U}$ denotes the uniform distribution and $N$ is the number of the required samples. This approach is summarized in Algorithm \ref{Algo3}. This sampling method is simple though, it is not necessarily the best. In particular, we should take the a priori information about the objective function into account as long as it is available. For example, if it is known that a minimum of the objective function $f$ is more likely to appear in a certain region, then the sample points should be more densely distributed in that region. 

\begin{algorithm}[t]
	\caption{Draw samples randomly}
	\label{Algo3}
	\hspace*{0.02in} {\bf Input:} $\bm{x}_L,\bm{x}_U$, and $N$
	\begin{algorithmic}[1]
		\For{$d=1,\cdots,D$}
		    \For{$n=1,\cdots,N$}
		    \State$x_n^d\leftarrow x_L^d+(x_U^d-x_L^d)r\quad\text{where}\quad r\sim\mathcal{U}([0,1])$
		    \EndFor
		\EndFor
	\end{algorithmic}
	\hspace*{0.02in} {\bf Output:} $\bm{x}_{n},n=1,\cdots,N$
\end{algorithm}

\subsection{Cut off a subregion from $\Omega$}\label{Sec2.2} 
In this section, we show how to cut off a subregion from $\Omega$ based on current observations. As before, we assume that $\Omega$ is a rectangular region for simplicity. In this case, a simple approach for cutting off a subregion from the rectangular region $\Omega$ is to reduce the size of each of its edge in a manner such that the current best sample point is close to the center of the reduced region as much as possible. Put it more specifically, given the rectangular feasible region $\Omega=[\bm{x}_L,\bm{x}_U]$ and the current minimum point $\bm{x}_{\min}$, we construct a new rectangular region $[\bm{x}'_L,\bm{x}'_U]$ with $\bm{x}_{\min}$ as its center and the size of each of its edge is reduced by a scalar $\lambda^n$ with respect to that of $\Omega$, where $\lambda<1$ is a parameter, i.e., we have
\begin{equation}\label{LU_Update}
\begin{split}
\bm{x}_L' &= \bm{x}_{\min}-\frac{\lambda^n}{2}(\bm{x}_U-\bm{x}_L)\\
\bm{x}_U' &= \bm{x}_{\min}+\frac{\lambda^n}{2}(\bm{x}_U-\bm{x}_L)\\
\end{split}
\end{equation}
If the new rectangular region $[\bm{x}'_L,\bm{x}'_U]$ is contained in the feasible region $\Omega$, then we are done and $\Omega_n\coloneqq[\bm{x}'_L,\bm{x}'_U]$. If not, we can move $[\bm{x}'_L,\bm{x}'_U]$ along each of its edge to make sure that it is exactly contained in the feasible region $\Omega$. This procedure is summarized in Algorithm \ref{Algo4}.

\begin{algorithm}[t]
	\caption{Cut off a subregion from $\Omega$}
	\label{Algo4}
	\hspace*{0.02in} {\bf Input:} $\Omega=[\bm{x}_L,\bm{x}_U], \bm{x}_{\min}$, and $\lambda$
	\begin{algorithmic}[1]
		\State calculate $\bm{x}_L'$ and $\bm{x}_U'$ by Eq. \eqref{LU_Update}
		\For{$d=1,\cdots,D$}
		    \If{$x_L'^d<x_L^d$}
		    \State $\quad\quad$$x_U'^d\leftarrow x_U'^d+x_L^d-x_L'^d,\quad x_L'^d\leftarrow x_L^d$
		    \ElsIf{$x_U'^d>x_U^d$}
		    \State $\quad\quad$$x_L'^d\leftarrow x_L'^d+x_U^d-x_U'^d,\quad x_U'^d\leftarrow x_U^d$
		    \EndIf
		\EndFor
	\end{algorithmic}
	\hspace*{0.02in} {\bf Output:} $\Omega_n=[\bm{x}_L',\bm{x}_u']$
\end{algorithm}

\subsection{A practical optimization by cut algorithm}\label{Sec2.3}

Put Algorithm \ref{Algo1}, \ref{Algo2} (or \ref{Algo3}), and \ref{Algo4} together, we obtain the following practical optimization by cut algorithm. It should be noticed that Algorithm \ref{Algo5} will terminates after a finite number of iterations, as can be seen from Eq. \eqref{LU_Update}. 

\begin{algorithm}[t]
	\caption{Practical Optimization by Cut}
	\label{Algo5}
	\hspace*{0.02in} {\bf Input:} $f:\mathbb{R}^D\to\mathbb{R},\Omega=[\bm{x}_L,\bm{x}_U],N,\lambda<1,\varepsilon>0$
	\begin{algorithmic}[1]
		\State $f_0\leftarrow+\infty, \bm{x}_0\leftarrow\bm{x}_L$, and $\Omega_0\leftarrow\Omega$
		\For{$n=1,2,\cdots$}
		\State draw $N$ samples $\bm{x}_1,\cdots,\bm{x}_N$ from $\Omega_{n-1}$ using Algorithm \ref{Algo2} or \ref{Algo3}
		\State $i\leftarrow\arg\min_{0\le i\le N}f(\bm{x}_i),\bm{x}_0\leftarrow\bm{x}_i,f_0\leftarrow f(\bm{x}_0)$
		\State construct $\Omega_n$ using Algorithm \ref{Algo4} with $\bm{x}_{\min}=\bm{x}_0$
		\State \textbf{quit} if $\max_{1\le d\le D}(x_U'^d-x_L'^d)<\varepsilon$
		\EndFor
	\end{algorithmic}
	\hspace*{0.02in} {\bf Output:} $\bm{x}_{\min}=\bm{x}_0$ and $f_{\min}=f_0$
\end{algorithm}

\section{Benchmark Functions}\label{Sec3}
To evaluate the performance of the proposed algorithm, we have tested it on most benchmark functions given in \cite{jamil2013literature}. However, due to limited space, it is unrealistic to present all the results here. Thus, we have selected $50$ of them which we hope are representative enough. Among these selected functions, $20$ are $2$-dimensional, $10$ are $4$-dimensional, and $20$ are $30$-dimensional. The functions given in \cite{jamil2013literature} contains many typos and the global minima given there are not comprehensive and accurate enough (or even wrong). In order to correct these errors and for ease of reference, we record the selected functions in Section \ref{Sec3.1}, \ref{Sec3.2}, and \ref{Sec3.3}, respectively. A more reliable source of these functions can be found in \cite{Al-Roomi2015}.

\subsection{2-dimensional cases}\label{Sec3.1}

\begin{enumerate}[1]
	\item \textbf{Ackley 3 Function \cite{jamil2013literature}:} This function is defined on $-32\le x_i\le 32$ by
	\begin{equation*}
	f_1(\bm{x}) = -200e^{-0.02\sqrt{x_1^2+x_2^2}}+5e^{\cos(3x_1)+\sin(3x_2)}
	\end{equation*}
	The global minimum is $f_1(\bm{x}^\ast)=-234.8853900346117$ and located at $\bm{x}^\ast=(0,0.511681300749165)$.
		
	\item \textbf{Beale Function \cite{jamil2013literature}:} This function is defined on $-4.5\le x_i\le 4.5$ by 
	\begin{equation*}
	f_2(\bm{x})=(1.5-x_1+x_1x_2)^2+(2.25-x_1+x_1x_2^2)^2+(2.625-x_1+x_1x_2^3)^2
	\end{equation*}
	The global minimum is $f_2(\bm{x}^\ast)=0$ and located at $\bm{x}^\ast=(3,0.5)$.
	
	\item \textbf{Booth Function \cite{jamil2013literature}:} This function is defined on $-10\le x_i\le 10$ by 
	\begin{equation*}
	f_3(\bm{x})=(x_1+2x_2-7)^2+(2x_1+x_2-5)^2
	\end{equation*}
	The global minimum is $f_3(\bm{x}^\ast)=0$ and located at $\bm{x}^\ast=(1,3)$.
	
	\item \textbf{Bukin 2 Function \cite{jamil2013literature}:} This function is defined on $[-15,-5]\times[-3,3]$ by 
	\begin{equation*}
	f_4(\bm{x})=100(x_2-0.01x_1^2+1)^2+0.01(x_1+10)^2
	\end{equation*}
	The global minimum is $f_2(\bm{x}^\ast)=0$ and located at $\bm{x}^\ast=(-10,0)$.
	
	\item \textbf{Camel Function-3 Hump \cite{jamil2013literature}:} This function is defined on $-5\le x_i\le 5$ by 
	\begin{equation*}
	f_5(\bm{x})=2x_1^2-1.05x_1^4+x_1^6/6+x_1x_2+x_2^2
	\end{equation*}
	The global minimum is $f_5(\bm{x}^\ast)=0$ and located at $\bm{x}^\ast=(0,0)$.
	
	\item \textbf{Chen Bird Function \cite{chen2003computer}:} This function is defined on $-500\le x_i\le 500$ by 
	\begin{equation*}
	f_6(\bm{x})=-\frac{b}{b^2+(x_1^2+x_2^2-1)^2}-\frac{b}{b^2+(x_1^2+x_2^2-\frac{1}{2})^2}-\frac{b}{b^2+(x_1-x_2)^2}
	\end{equation*}
	where $b=0.001$. It has four global minima $f_6(\bm{x}^\ast)=-2000.003999984001$ that are located at $\bm{x}^\ast=\pm(0.5,0.5)$ and $\pm(\sqrt{2}/2,\sqrt{2}/2)$.
	
	\item \textbf{Cube Function \cite{jamil2013literature}:} This function is defined on $-10\le x_i\le 10$ by
	\begin{equation*}
	f_7(\bm{x})=100(x_2-x_1^3)^2+(1-x_1)^2
	\end{equation*}
	The global minimum is $f_7(\bm{x}^\ast)=0$ and located at $\bm{x}^\ast=(1,1)$.
	
	\item \textbf{Damavandi Function \cite{jamil2013literature}:} This function is defined on $0\le x_i\le 14$ by
	\begin{equation*}
	f_8(\bm{x})=\left(1-\left|\frac{\sin[\pi(x_1-2)]\sin[\pi(x_2-2)]}{\pi^2(x_1-2)(x_2-2)}\right|^5\right)\left(2+(x_1-7)^2+2(x_2-7)^2\right)
	\end{equation*}
	The global minimum is $f_8(\bm{x}^\ast)=0$ and located at $\bm{x}^\ast=(2,2)$.
	
	\item \textbf{Jennrich-Sampson Function \cite{jamil2013literature}:} This function is defined on $-1\le x_i\le 1$ by
	\begin{equation*}
	f_9(\bm{x})=\sum_{i=1}^{10}(2+2i-(e^{ix_1}+e^{ix_2}))^2
	\end{equation*}
	The global minimum is $f_9(\bm{x}^\ast)=124.36218235561473896$ and located at $\bm{x}^\ast=(0.257825214197515,0.257825213363251)$.
	
	\item \textbf{Leon Function \cite{jamil2013literature}:} This function is defined on $-1.2\le x_i\le 1.2$ by
	\begin{equation*}
	f_{10}(\bm{x})=100(x_2-x_1^2)^2+(1-x_1)^2
	\end{equation*}
	The global minimum is $f_{10}(\bm{x}^\ast)=0$ and located at $\bm{x}^\ast=(1,1)$.
		
	\item \textbf{Matyas Function \cite{jamil2013literature}:} This function is defined on $-10\le x_i\le 10$ by
	\begin{equation*}
	f_{11}(\bm{x})=0.26(x_1^2+x_2^2)-0.48x_1x_2
	\end{equation*}
	The global minimum is $f_{11}(\bm{x}^\ast)=0$ and located at $\bm{x}^\ast=(0,0)$.
	
	\item \textbf{Mishra 3 Function \cite{Al-Roomi2015}:} This function is defined on $-10\le x_i\le 10$ by
	\begin{equation*}
	f_{12}(\bm{x})=\sqrt{\left|\cos\sqrt{\left|x_1^2+x_2\right|}\right|}+0.01(x_1+x_2)
	\end{equation*}
	The global minimum is $f_{12}(\bm{x}^\ast)=-0.184666993496657$ and located at $\bm{x}^\ast=(-8.466701099413424,-10)$.
	
	\item \textbf{Mishra 10 a Function \cite{Al-Roomi2015}:} This function is defined on $-10\le x_i\le 10$ by
	\begin{equation*}
	f_{13}(\bm{x})=\left(x_1+x_2-x_1x_2\right)^2
	\end{equation*}
	It has two global minima  $f_{13}(\bm{x}^\ast)=0$ located at $\bm{x}^\ast=(0,0)$ and $(2,2)$.
	
	\item \textbf{Price 2 Function \cite{jamil2013literature}:} This function is defined on $-10\le x_i\le 10$ by
	\begin{equation*}
	f_{14}(\bm{x})=1+\sin^2x_1+\sin^2x_2-0.1e^{-x_1^2-x_2^2}
	\end{equation*}
	The global minimum is $f_{14}(\bm{x}^\ast)=0.9$ and located at $\bm{x}^\ast=(0,0)$.
	
	\item \textbf{Scahffer 1 Function \cite{jamil2013literature}:} This function is defined on $-100\le x_i\le 100$ by
	\begin{equation*}
	f_{15}(\bm{x})=0.5+\frac{\sin^2(x_1^2+x_2^2)^2-0.5}{1+0.001(x_1^2+x_2^2)^2}
	\end{equation*}
	The global minimum is $f_{15}(\bm{x}^\ast)=0$ and located at $\bm{x}^\ast=(0,0)$.
	
	\item \textbf{Schwefel 2.6 Function \cite{jamil2013literature}:} This function is defined on $-100\le x_i\le 100$ by
	\begin{equation*}
	f_{16}(\bm{x})=\max(|x_1+2x_2-7|,|2x_1+x_2-5|)
	\end{equation*}
	The global minimum is $f_{16}(\bm{x}^\ast)=0$ and located at $\bm{x}^\ast=(1,3)$.
	
	\item \textbf{Testtube Holder Function \cite{Al-Roomi2015}:} This function is defined on $-10\le x_i\le 10$ by
	\begin{equation*}
	f_{17}(\bm{x})=-4\left|\sin x_1\cos x_2 e^{|\cos((x_1^2+x_2^2)/200)|}\right|
	\end{equation*}
	It has two global minima $f_{17}(\bm{x}^\ast)= -10.872300105622747$ that are located at $\bm{x}^\ast=(\pm1.570602622190189,0)$.
	
	\item \textbf{Trefethen Function \cite{jamil2013literature}:} This function is defined on $-10\le x_i\le 10$ by
	\begin{multline*}
	f_{18}(\bm{x})=e^{\sin(50x_1)}+\sin(60e^{x_2})+\sin(70\sin(x_1))\\
	+\sin(\sin(80x_2))-\sin(10(x_1+x_2))+\frac{1}{4}(x_1^2+x_2^2)
	\end{multline*}
	The global minimum is $f_{18}(\bm{x}^\ast)=-3.306868647475237$ and located at $\bm{x}^\ast=(-0.024403079433617, 0.210612427428984)$.
	
	\item \textbf{Tripod Function \cite{Al-Roomi2015}:} This function is defined on $-100\le x_i\le 100$ by
	\begin{equation*}
	f_{19}(\bm{x})=p_2(1+p_1)+\left|x_1+50p_2(1-2p_1)\right|+|x_2+50(1-2p_2)|
	\end{equation*}
	where 
	\begin{equation*}
	p_i=\begin{cases}
	1&\text{if }x_i\ge 0\\
	0&\text{otherwise}
	\end{cases}
	\end{equation*}
	The global minimum is $f_{19}(\bm{x}^\ast)=0$ and located at $\bm{x}^\ast=(0,-50)$.
	
	\item \textbf{Wayburn Seader 2 Function \cite{Al-Roomi2015}:} This function is defined on $-500\le x_i\le 500$ by
	\begin{equation*}
	f_{20}(\bm{x})=\left(1.613-4(x_1-0.3125)^2-4(x_2-1.625)^2\right)^2+(x_2-1)^2
	\end{equation*}
	It has two global minima $f_{20}(\bm{x}^\ast)=0$ located at $\bm{x}^\ast=(0.424861025271221, 1)$ and $(0.200138974728779, 1)$ (or in analytic form as $\bm{x}^\ast=(0.3125\pm\frac{\sqrt{0.0505}}{2}, 1)$).
\end{enumerate}

\subsection{4-dimensional cases}\label{Sec3.2}

\begin{enumerate}[1]
	\item \textbf{Biggs EXP4 Function \cite{biggs1971minimization}:} This function is defined on $0\le x_i\le 20$ by
	\begin{equation*}
	f_{21}(\bm{x})=\sum_{i=1}^{10}\left(x_3e^{-t_ix_1}-x_4e^{-t_ix_2}-e^{-t_i}+5e^{-10t_i}\right)^2
	\end{equation*}
	where $t_i=0.1i$. The global minimum is $f_{21}(\bm{x}^\ast)=0$ and located at $\bm{x}^\ast=(1, 10, 1, 5)$.
	
	\item \textbf{Colville Function \cite{jamil2013literature}:} This function is defined on $-10\le x_i\le 10$ by
	\begin{multline*}
	f_{22}(\bm{x})=100(x_1-x_2^2)^2+(1-x_1)^2+90(x_4-x_3^2)^2+(1-x_3)^2\\
	+10.1((x_2-1)^2+(x_4-1)^2)+19.8(x_2-1)(x_4-1)
	\end{multline*}
	The global minimum is $f_{22}(\bm{x}^\ast)=0$ and located at $\bm{x}^\ast=(1, 1, 1, 1)$.
	
	\item \textbf{Corana Function \cite{Al-Roomi2015}:} This function is defined on $-500\le x_i\le 500$ by
	\begin{equation*}
	f_{23}(\bm{x})=\sum_{i=1}^4\begin{cases}
	0.15d_i\left(z_i-0.05\mathrm{sign}(z_i)\right)^2&\text{if }\vert x_i-z_i\vert<0.05\\
	d_ix_i^2&\text{otherwise}
	\end{cases}
	\end{equation*}
	where 
	\begin{align*}
	z_i&=0.2\left\lfloor\left\vert\frac{x_i}{0.2}\right\vert+0.49999\right\rfloor\mathrm{sign}(x_i)\\
	d_i&=(1,1000,10,100)
	\end{align*}
	It has an infinite number of global minima $f_{23}(\bm{x}^\ast)=0$ with $\vert x_i^\ast\vert<0.05$.
	
	\item \textbf{deVilliers Glasser 1 Function \cite{jamil2013literature}:} This function is defined on $1\le x_i\le 100$ by
	\begin{equation*}
	f_{24}(\bm{x})=\sum_{i=1}^{24}\left(x_1x_2^{t_i}\sin(x_3t_i+x_4)-ab^{t_i}\sin(ct_i+d)\right)^2
	\end{equation*}
	where $t_i=0.1(i-1)$ and $(a,b,c,d)=(60.137,1.371,3.112,1.761)$. It has many global minima $f_{24}(\bm{x}^\ast)=0$ with $x_1^\ast=a,x_2^\ast=b$ and $x_3^\ast,x_4^\ast$ satisfy $\sin(x_3^\ast t_i+x_4^\ast)=\sin(c t_i+d),i=1,\cdots,24$.
	
	\item \textbf{Gear Function \cite{Al-Roomi2015}:} This function is defined on $12\le x_i\le 60$ by
	\begin{equation*}
	f_{25}(\bm{x})=\left(\frac{10}{6.931}-\frac{\lfloor x_1\rfloor\lfloor x_2\rfloor}{\lfloor x_3\rfloor\lfloor x_4\rfloor}\right)^2
	\end{equation*}
	It has an infinite number of global minima $f_{25}(\bm{x}^\ast)=2.700857148886513\times10^{-12}$ with $\lfloor\bm{x}^\ast\rfloor=(16, 19, 43, 49)$ (the value of $x_1$ and $x_2$ or $x_3$ and $x_4$ can be permuted).
	
	\item \textbf{Miele Cantrell Function \cite{jamil2013literature}:} This function is defined on $-1\le x_i\le 1$ by
	\begin{equation*}
	f_{26}(\bm{x})=(e^{-x_1}-x_2)^4+100(x_2-x_3)^6+\tan^4(x_3-x_4)+x_1^8
	\end{equation*}
	The global minimum is $f_{26}(\bm{x}^\ast)=0$ and located at $\bm{x}^\ast=(0, 1, 1, 1)$.
	
	\item \textbf{Powell Singular Function \cite{Al-Roomi2015}:} This function is defined on $-4\le x_i\le 5$ by
	\begin{equation*}
	f_{27}(\bm{x})=(x_1+10x_2)^2+5(x_3-x_4)^2+(x_2-x_3)^4+10(x_1-x_4)^4
	\end{equation*}
	The global minimum is $f_{27}(\bm{x}^\ast)=0$ and located at $\bm{x}^\ast=(0,0,0,0)$.
	
	\item \textbf{Shekel 5 Function \cite{jamil2013literature}:} This function is defined on $0\le x_i\le 10$ by
	\begin{equation}\label{Shekel}
	f_{28}(\bm{x})=-\sum_{i=1}^n\frac{1}{\sum_{j=1}^4(x_j-a_{ij})^2+c_i}
	\end{equation}
	with $n=5$, where 
	\begin{equation*}
	[a_{ij}]=\begin{bmatrix}
	4&1&8&6&3&2&5&8&6&7\\
	4&1&8&6&7&9&3&1&2&3.6\\
	4&1&8&6&3&2&5&8&6&7\\
	4&1&8&6&7&9&3&1&2&3.6\\
	\end{bmatrix}^T
	\end{equation*}
	and
	\begin{equation*}
	[c_i]=\begin{bmatrix}
	0.1&0.2&0.2&0.4&0.4&0.6&0.3&0.7&0.5&0.5
	\end{bmatrix}^T
	\end{equation*}
	The global minimum is $f_{28}(\bm{x}^\ast)=-10.153199679058231$ and located at $x_1^\ast=x_3^\ast=4.000037152015988$ and $x_2^\ast=x_4^\ast=4.000133277358568$.
	
	\item \textbf{Shekel 7 Function \cite{jamil2013literature}:} This function is defined on $0\le x_i\le 10$ by \eqref{Shekel} with $n=7$. The global minimum is $f_{29}(\bm{x}^\ast)=-10.402915336777747$ and located at $x_1^\ast=x_3^\ast=4.000572820035435$ and $x_2^\ast=x_4^\ast=3.999606208991378$.
	
	\item \textbf{Shekel 10 Function \cite{jamil2013literature}:} This function is defined on $0\le x_i\le 10$ by \eqref{Shekel} with $n=10$. The global minimum is $f_{30}(\bm{x}^\ast)=-10.536443153483534$ and located at $x_1^\ast=x_3^\ast=4.000746868833048$ and $x_2^\ast=x_4^\ast=3.999509479273299$.

\end{enumerate}

\subsection{30-dimensional cases}\label{Sec3.3}

\begin{enumerate}[1]
	\item \textbf{Ackely 1 Function \cite{jamil2013literature}:} This function is defined on $-32\le x_i\le 32$ by
	\begin{equation*}
	f_{31}(x)=-20e^{-0.02\sqrt{D^{-1}\sum_{i=1}^Dx_i^2}}-e^{D^{-1}\sum_{i=1}^D\cos(2\pi x_i)}+20+e
	\end{equation*}
	The global minimum is $f_{31}(\bm{x}^\ast)=0$ and located at $\bm{x}^\ast=(0,\cdots,0)$.
	
	\item \textbf{Cosine Mixture Function \cite{jamil2013literature}:} This function is defined on $-1\le x_i\le 1$ by
	\begin{equation*}
	f_{32}(x)=-0.1\sum_{i=1}^D\cos(5\pi x_i)+\sum_{i=1}^Dx_i^2
	\end{equation*}
	The global minimum is $f_{32}(\bm{x}^\ast)=-0.1D$ and located at $\bm{x}^\ast=(0,\cdots,0)$.
	
	\item \textbf{Csendes Function \cite{jamil2013literature}:} This function is defined on $-1\le x_i\le 1$ by
	\begin{equation*}
	f_{33}(x)=\sum_{i=1}^Dx_i^6\left(2+\sin\frac{1}{x_i}\right)
	\end{equation*}
	The global minimum is $f_{33}(\bm{x}^\ast)=0$ and located at $\bm{x}^\ast=(0,\cdots,0)$.
	
	\item \textbf{Deb 1 Function \cite{jamil2013literature}:} This function is defined on $-1\le x_i\le 1$ by
	\begin{equation*}
	f_{34}(x)=-\frac{1}{D}\sum_{i=1}^D\sin^6(5\pi x_i)
	\end{equation*}
	It has $10^D$ global minima $f_{34}(\bm{x}^\ast)=-1$ with $x_i^\ast=\pm0.1,\pm0.3\cdots,\pm0.9$.
	
	\item \textbf{Dixon \& Price Function \cite{jamil2013literature}:} This function is defined on $-10\le x_i\le 10$ by
	\begin{equation*}
	f_{35}(x)=(x_1-1)^2+\sum_{i=2}^Di(2x_i^2-x_{i-1})^2
	\end{equation*}
	The global minimum is $f_{35}(\bm{x}^\ast)=0$ and located at $x_i^\ast= 2^{(1/2^{(i-1)}-1)}$.
	
	\item \textbf{Exponential Function \cite{jamil2013literature}:} This function is defined on $-1\le x_i\le 1$ by
	\begin{equation*}
	f_{36}(x)=-\exp\left(-\frac{1}{2}\sum_{i=1}^Dx_i^2\right)
	\end{equation*}
	The global minimum is $f_{36}(\bm{x}^\ast)=-1$ and located at $\bm{x}^\ast=(0,\cdots,0)$.
	
	\item \textbf{Griewank Function \cite{jamil2013literature}:} This function is defined on $-100\le x_i\le 100$ by
	\begin{equation*}
	f_{37}(x)=\sum_{i=1}^D\frac{x_i^2}{4000}-\prod_{i=1}^D\cos\left(\frac{x_i}{\sqrt{i}}\right)+1
	\end{equation*}
	The global minimum is $f_{33}(\bm{x}^\ast)=0$ and located at $\bm{x}^\ast=(0,\cdots,0)$.
	
	\item \textbf{Mishra 1 Function \cite{jamil2013literature}:} This function is defined on $0\le x_i\le 1$ by
	\begin{equation*}
	f_{38}(x)=\left(1+g\right)^g
	\end{equation*}
	where $g=1+D-\sum_{i=1}^Dx_i$. The global minimum is $f_{38}(\bm{x}^\ast)=2$ and located at $\bm{x}^\ast=(1,\cdots,1)$.
	
	\item \textbf{Powell Sum Function \cite{jamil2013literature}:} This function is defined on $-1\le x_i\le 1$ by
	\begin{equation*}
	f_{39}(x)=\sum_{i=1}^D\left|x_i\right|^{i+1}
	\end{equation*}
	The global minimum is $f_{39}(\bm{x}^\ast)=0$ and located at $\bm{x}^\ast=(0,\cdots,0)$.
	
	\item \textbf{Qing Function \cite{jamil2013literature}:} This function is defined on $-500\le x_i\le 500$ by
	\begin{equation*}
	f_{40}(x)=\sum_{i=1}^D(x_i^2-i)^2
	\end{equation*}
	It has $2^D$ global minima $f_{40}(\bm{x}^\ast)=0$ with $x_i^\ast=\pm\sqrt{i}$.
		
	\item \textbf{Quintic Function \cite{jamil2013literature}:} This function is defined on $-10\le x_i\le 10$ by
	\begin{equation*}
	f_{41}(x)=\sum_{i=1}^D\left|x_i^5-3x_i^4+4x_i^3+2x_i^2-10x_i-4\right|
	\end{equation*}
	It has $3^D$ global minima $f_{41}(\bm{x}^\ast)=0$ with $x_i^\ast=-1,2$ or $-0.402627941186124$.
		
	\item \textbf{Rosenbrock Function \cite{jamil2013literature}:} This function is defined on $-30\le x_i\le 30$ by
	\begin{equation*}
	f_{42}(x)=\sum_{i=1}^{D-1}\left[100(x_{i+1}-x_i^2)^2+(x_i-1)^2\right]
	\end{equation*}
	The global minimum is $f_{42}(\bm{x}^\ast)=0$ and located at $\bm{x}^\ast=(1,\cdots,1)$.
	
	\item \textbf{Salomon Function \cite{jamil2013literature}:} This function is defined on $-100\le x_i\le 100$ by
	\begin{equation*}
	f_{43}(x)=1-\cos\left(2\pi\sqrt{\sum_{i=1}^Dx_i^2}\right)+0.1\sqrt{\sum_{i=1}^Dx_i^2}
	\end{equation*}
	The global minimum is $f_{43}(\bm{x}^\ast)=0$ and located at $\bm{x}^\ast=(0,\cdots,0)$.
	
	\item \textbf{Schwefel Function \cite{jamil2013literature}:} This function is defined on $-100\le x_i\le 100$ by
	\begin{equation*}
	f_{44}(x)=\left(\sum_{i=1}^Dx_i^2\right)^\alpha
	\end{equation*}
	where $\alpha=0.1$. The global minimum is $f_{44}(\bm{x}^\ast)=0$ and located at $\bm{x}^\ast=(0,\cdots,0)$.
	
	\item \textbf{Stepint Function \cite{Al-Roomi2015}:} This function is defined on $-5.12\le x_i\le 5.12$ by
	\begin{equation*}
	f_{45}(x)=25+\sum_{i=1}^D\lfloor x_i\rfloor
	\end{equation*}
	It has an infinite number of minima $f_{45}(\bm{x}^\ast)=25-6D$ with $x_i^\ast\in[-5.12,-5)$.
	
	\item \textbf{Streched V Sine Wave Function \cite{jamil2013literature}:} This function is defined on $-10\le x_i\le 10$ by
	\begin{equation*}
	f_{46}(x)=\sum_{i=1}^{D-1}(x_{i+1}^2+x_i^2)^{0.25}\left(\sin^2(50(x_{i+1}^2+x_i^2)^{0.1})+0.1\right)
	\end{equation*}
	The global minimum is $f_{46}(\bm{x}^\ast)=0$ and located at $\bm{x}^\ast=(0,\cdots,0)$.
	
	\item \textbf{W/Wavy Function \cite{jamil2013literature}:} This function is defined on $-\pi\le x_i\le\pi$ by
	\begin{equation*}
	f_{47}(x)=1-\frac{1}{D}\sum_{i=1}^D\cos(kx_i)e^{\frac{-x_i^2}{2}}
	\end{equation*}
	where $k=10$. The global minimum is $f_{47}(\bm{x}^\ast)=0$ and located at $\bm{x}^\ast=(0,\cdots,0)$.
	
	\item \textbf{Weierstrass Function \cite{suganthan2005problem}:} This function is defined on $-0.5\le x_i\le 0.5$ by
	\begin{equation*}
	f_{48}(x)=\sum_{i=1}^D\sum_{k=0}^{K}a^k\left(\cos(2\pi b^k(x_i+0.5))-\cos(\pi b^k)\right)
	\end{equation*}
	where $a=0.5,b=3,$ and $K=20$. The global minimum is $f_{48}(\bm{x}^\ast)=0$ and located at $\bm{x}^\ast=(0,\cdots,0)$.
	
	\item \textbf{Whitley Function \cite{jumonji2007novel}:} This function is defined on $-10\le x_i\le 10$ by
	\begin{equation*}
	f_{49}(x)=\sum_{i=1}^D\sum_{j=1}^D\left(\frac{g_{ij}^2}{4000}-\cos g_{ij}+1\right)
	\end{equation*}
	where $g_{ij}=100(x_i^2-x_j)^2+(1-x_j)^2$. The global minimum is $f_{49}(\bm{x}^\ast)=0$ and located at $\bm{x}^\ast=(1,\cdots,1)$.
	
	\item \textbf{Zakharov Function \cite{jamil2013literature}:} This function is defined on $-5\le x_i\le 10$ by
	\begin{equation*}
	f_{50}(x)=\sum_{i=1}^Dx_i^2+\left(\frac{1}{2}\sum_{i=1}^Dix_i\right)^2+\left(\frac{1}{2}\sum_{i=1}^Dix_i\right)^4
	\end{equation*}
	The global minimum is located at $\bm{x}^\ast=(0,\cdots,0)$ with $f_{50}(\bm{x}^\ast)=0$.
	
\end{enumerate}

\section{Performance Evaluation of the Proposed Algorithms}\label{Sec4}

In this section, the performance of the proposed algorithms on the selected benchmark functions given in Section \ref{Sec3} is compared with that of the PSO. To make the results more convincing, we have implement $100$ independent runs for each objective function and each algorithm. We use the median of the error $\epsilon=f-f(\bm{x}^\ast)$ and the average of the computational time $t$ as performance metrics. The reason why the median rather than the average (as proposed by \cite{bratton2007defining}) of the error $\epsilon$ is used can be explained as follows. If an algorithm performs very well in $99$ runs with a very low error, say $10^{-9}$, but very bad in one single run with an error of $1$, then the average error would be  approximately $1$, which is apparently a bad representative of the performance of the algorithm.

\subsection{2-dimensional case}\label{Sec4.1}

In the $2$-dimensional case, we have evaluated the performance of Algorithm \ref{Algo5} with both the deterministic (Algorithm \ref{Algo2}) and the stochastic (Algorithm \ref{Algo3}) sampling strategies. These two schemes are referred to, respectively, as OCD (Optimization by Cut with deterministic sampling strategy) and OCS (Optimization by Cut with stochastic sampling strategy) in the rest of the paper. The results are compared to that of the PSO given by Eq. \eqref{PSO_Update}. For both the OCD and OCS schemes, the number of iterations is set to $50$ and the parameter $\lambda$ is set to $0.4$. The number of samples per iteration is $N^D=30^2=900$, which corresponds to a total of $50\times 900=45000$ evaluations of each objective function. For the PSO scheme, the parameters $w=0.4$ and $c_1=c_2=1.5$. Both the number of iterations and the number of particles are set to $100$ which corresponds to a total of $100\times 100=10000$ evaluations of each objective function. 

\begin{figure}[htbp]
	\centering
	\includegraphics[width=\linewidth]{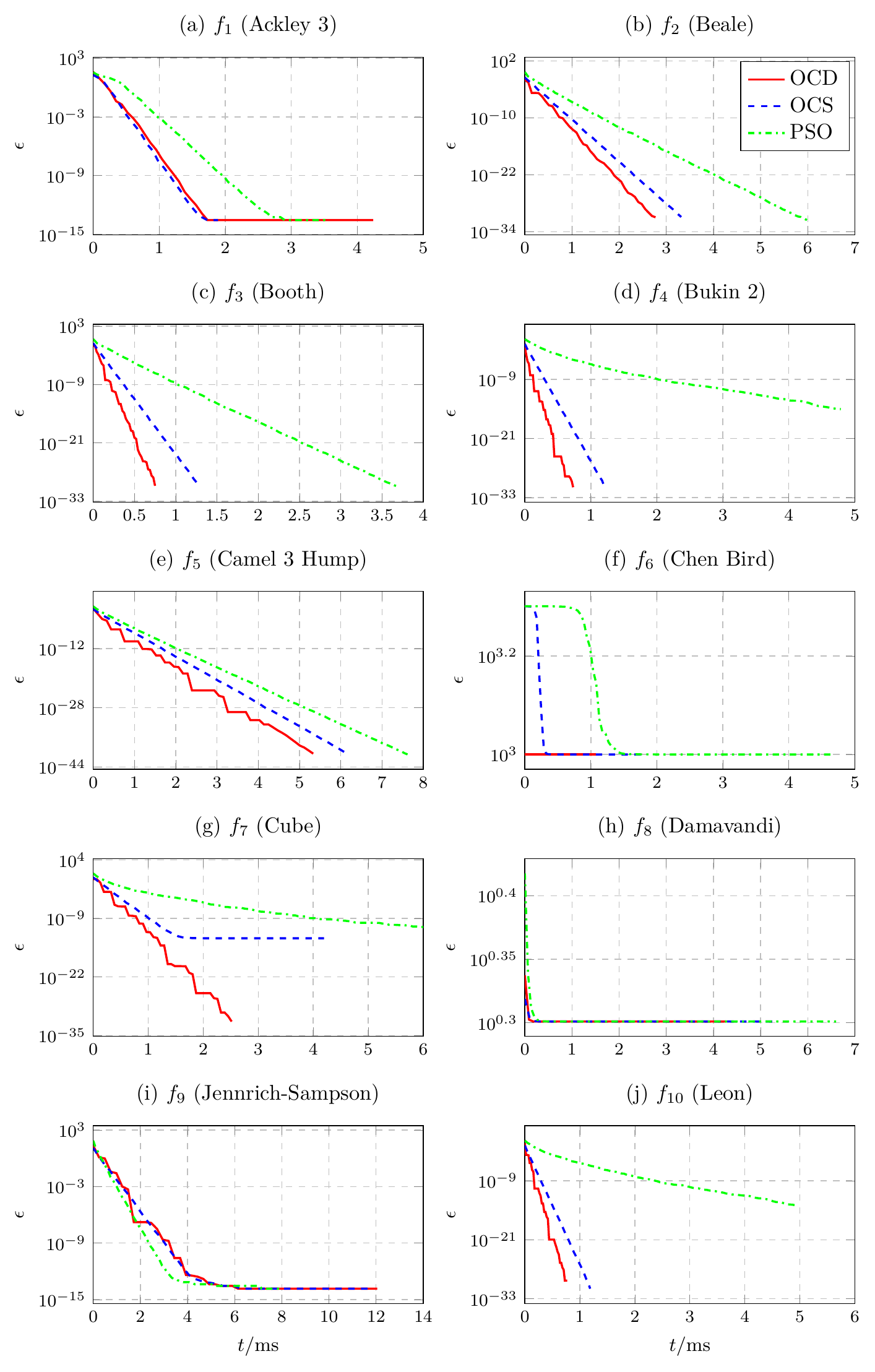}
	\caption{2-dimensional functions}
	\label{Fig_2d_1}
\end{figure}

\begin{figure}[htbp]
	\centering
	\includegraphics[width=\linewidth]{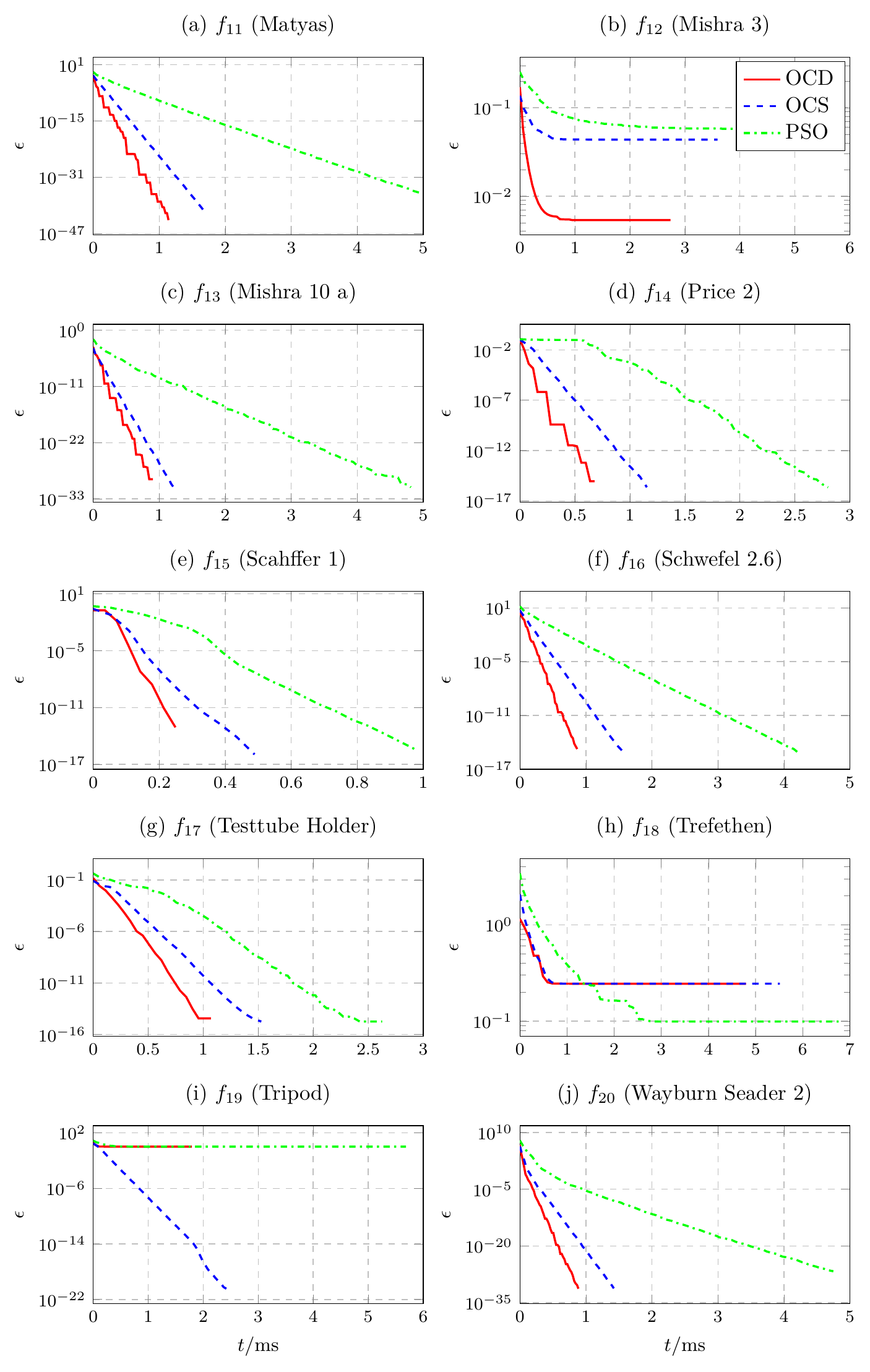}
	\caption{2-dimensional functions}
	\label{Fig_2d_2}
\end{figure}

\begin{table}[htbp]
	\caption{Final error and computational time ($2$-dimensional cases)}
	\label{Tab_2d_1}
	\centering
	\begin{tabular}{llll|lll}
		\hline
		&    &             $\epsilon$ &    $t/s$ &                           &             $\epsilon$ &    $t/s$\\
		\hline
		\multirow{3}{*}{$f_1$}  &OCD & $2.8422\times10^{-14}$ & $0.0042$ &    \multirow{3}{*}{$f_6$} &            $1000.0040$ & $0.0011$\\
								&OCS &                    $0$ & $0.0040$ &                           &            $1000.0040$ & $0.0018$\\
								&PSO &                    $0$ & $0.0065$ &                           &            $1000.0040$ & $0.0047$\\
		\hline
		\multirow{3}{*}{$f_2$}  &OCD &                    $0$ & $0.0036$ &    \multirow{3}{*}{$f_7$} &                    $0$ & $0.0032$\\
								&OCS &                    $0$ & $0.0044$ &                           & $3.9226\times10^{-14}$ & $0.0042$\\
								&PSO &                    $0$ & $0.0064$ &	                         & $1.0660\times10^{-11}$ & $0.0061$\\
		\hline
		\multirow{3}{*}{$f_3$}  &OCD &                    $0$ & $0.0010$ &    \multirow{3}{*}{$f_8$} &                    $2$ & $0.0043$\\
								&OCS &                    $0$ & $0.0017$ &                           &                    $2$ & $0.0053$\\
								&PSO &                    $0$ & $0.0047$ &                           &                    $2$ & $0.0066$\\
		\hline
		\multirow{3}{*}{$f_4$}  &OCD &                    $0$ & $0.0010$ &    \multirow{3}{*}{$f_9$} & $1.4211\times10^{-14}$ & $0.0120$\\
								&OCS &                    $0$ & $0.0016$ &                           & $1.4211\times10^{-14}$ & $0.0117$\\
								&PSO & $9.0709\times10^{-16}$ & $0.0048$ &                           & $1.4211\times10^{-14}$ & $0.0080$\\
		\hline
		\multirow{3}{*}{$f_5$}  &OCD & $4.0486\times10^{-41}$ & $0.0053$ & \multirow{3}{*}{$f_{10}$} &                    $0$ & $0.0011$\\
								&OCS & $2.6048\times10^{-41}$ & $0.0062$ &                           &                    $0$ & $0.0016$\\
								&PSO & $2.5274\times10^{-41}$ & $0.0076$ &                           & $8.3655\times10^{-15}$ & $0.0050$\\
		\hline
	\end{tabular}
\end{table}

\begin{table}[htbp]
	\caption{Final error and computational time ($2$-dimensional cases)}
	\label{Tab_2d_2}
	\centering
	\begin{tabular}{llll|lll}
		\hline
		&    &             $\epsilon$ &    $t/s$ &                           &             $\epsilon$ &    $t/s$\\
		\hline
		\multirow{3}{*}{$f_{11}$}  &OCD & $6.2640\times10^{-44}$ & $0.0011$ & \multirow{3}{*}{$f_{16}$} &                    $0$ & $0.0011$\\
		&OCS & $7.8040\times10^{-42}$ & $0.0017$ &                           &                    $0$ & $0.0019$\\
		&PSO & $4.1578\times10^{-36}$ & $0.0049$ &                           &                    $0$ & $0.0049$\\
		\hline
		\multirow{3}{*}{$f_{12}$}  &OCD &               $0.0054$ & $0.0027$ & \multirow{3}{*}{$f_{17}$} &                    $0$ & $0.0028$\\
		&OCS &               $0.0436$ & $0.0036$ &                           &                    $0$ & $0.0037$\\
		&PSO &               $0.0565$ & $0.0052$ &                           &                    $0$ & $0.0050$\\
		\hline
		\multirow{3}{*}{$f_{13}$}  &OCD &                    $0$ & $0.0014$ & \multirow{3}{*}{$f_{18}$} &               $0.2442$ & $0.0048$\\
		&OCS &                    $0$ & $0.0018$ &                           &               $0.2442$ & $0.0055$\\
		&PSO & $1.8527\times10^{-31}$ & $0.0048$ &                           &               $0.0987$ & $0.0068$\\
		\hline
		\multirow{3}{*}{$f_{14}$}  &OCD &                    $0$ & $0.0020$ & \multirow{3}{*}{$f_{19}$} &                    $1$ & $0.0018$\\
		&OCS &                    $0$ & $0.0030$ &                           & $3.5555\times10^{-21}$ & $0.0024$\\
		&PSO &                    $0$ & $0.0052$ &                           &                    $1$ & $0.0057$\\
		\hline
		\multirow{3}{*}{$f_{15}$}  &OCD &                    $0$ & $0.0017$ & \multirow{3}{*}{$f_{20}$} &                    $0$ & $0.0010$\\
		&OCS &                    $0$ & $0.0027$ &                           &                    $0$ & $0.0017$\\
		&PSO &                    $0$ & $0.0048$ &                           & $2.0474\times10^{-27}$ & $0.0047$\\
		\hline
	\end{tabular}
\end{table}

The evolution of the median error $\epsilon=f-f(\bm{x}^\ast)$ with respect to the average computational time $t$ for each objective function is shown in Fig. \ref{Fig_2d_1} and \ref{Fig_2d_2}. Moreover, the final error and computational time are collected in Tab. \ref{Tab_2d_1} and \ref{Tab_2d_2} for ease of comparison. From Fig. \ref{Fig_2d_1} and \ref{Fig_2d_2} we see that for most objective functions, the proposed OCD and OCS schemes converge to a global minimum faster than the PSO scheme and the OCD scheme is the best in terms of convergence rate. Moreover, from Tab. \ref{Tab_2d_1} and \ref{Tab_2d_2} we see that for most objective functions, the proposed OCD and OCS schemes consume less computational time than the PSO scheme even though they require more evaluations and have higher accuracy. From our perspective, this advantage stem from the simplicity of the proposed schemes. Specifically, they only need to update, in each iteration, the upper and lower bounds of a rectangular region by Eq. \ref{LU_Update}, which is computationally much simpler than the updating rules \ref{PSO_Update} of the PSO.

\subsection{4-dimensional case}\label{Sec4.2}

In the $4$-dimensional case, the proposed OCD and OCS schemes are also compared with the PSO scheme. For both the OCD and OCS schemes, the number of iterations is set to $200$ and the parameter $\lambda$ is set to $0.8$. The number of samples per iteration is $N^D=8^4=4096$ for the OCD scheme and $N=2000$ for the OCS scheme, which correspond to a total of $200\times 4096=819200$ and $200\times 2000=400000$ evaluations of each objective function, respectively. For the PSO scheme, the parameters $w=0.5$ and $c_1=c_2=1.5$. The number of iterations is set to $200$ and the number of particles is set to $1000$ which correspond to a total of $200\times 1000=200000$ evaluations of each objective function.  

\begin{figure}[htbp]
	\centering
	\includegraphics[width=\linewidth]{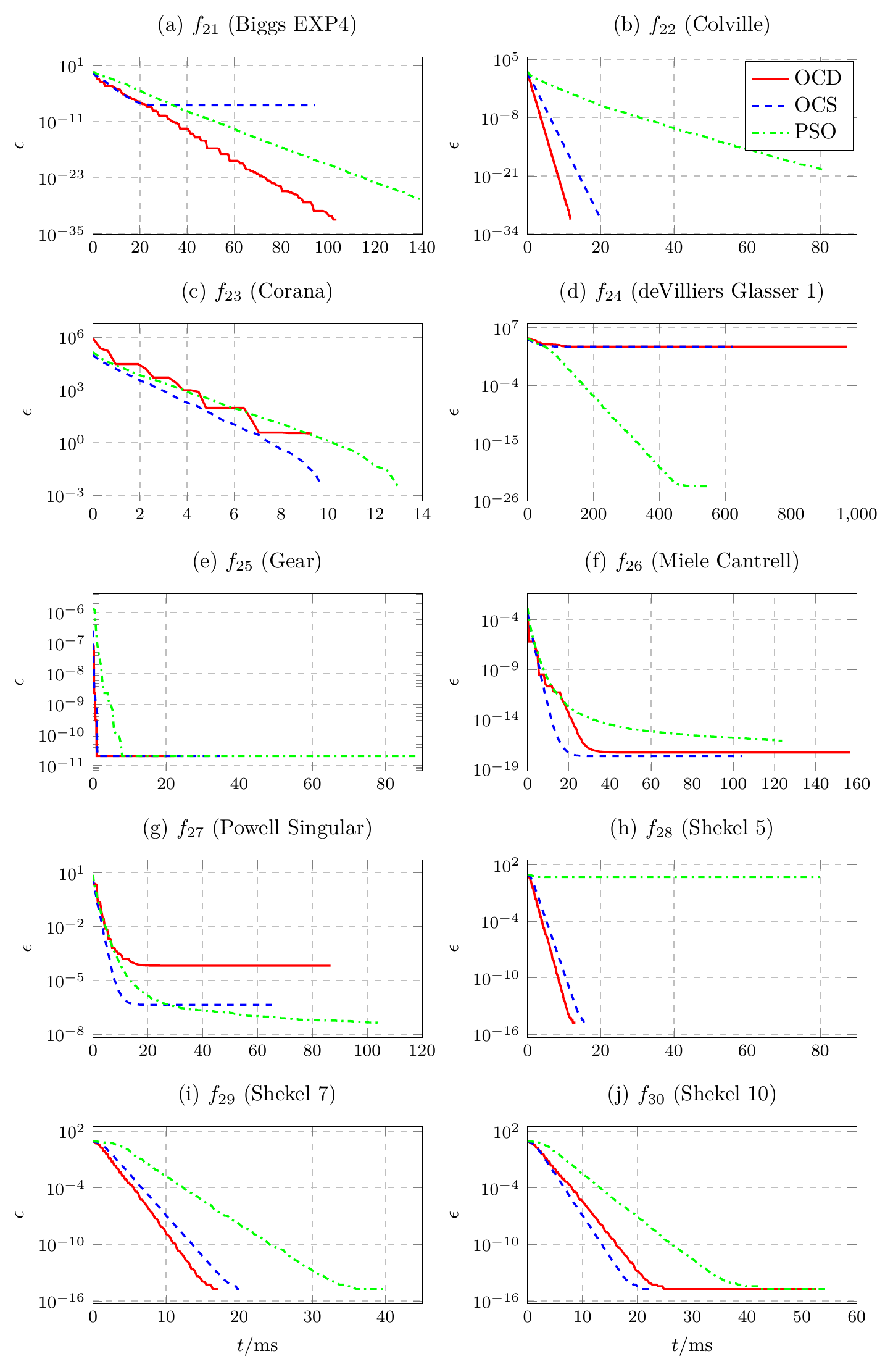}
	\caption{4-dimensional functions}
	\label{Fig_4d}
\end{figure}

The evolution of the error $\epsilon$ with respect to the computational time $t$ is presented in Fig. \ref{Fig_4d}. The final error and the computational time are collected in Tab. \ref{Tab_4d} for ease of comparison. In this case, we can draw a roughly the same conclusion as the $2$-dimensional case, so we are not going to repeat it here.

\begin{table}[htbp]
	\caption{Final error and computational time ($4$-dimensional cases)}
	\label{Tab_4d}
	\centering
	\begin{tabular}{llll|lll}
		\hline
		                           &    &             $\epsilon$ &    $t/s$ &                            &             $\epsilon$ &    $t/s$\\
		\hline
		\multirow{3}{*}{$f_{21}$}  &OCD &                    $0$ & $0.1272$ &  \multirow{3}{*}{$f_{26}$} & $4.4867\times10^{-18}$ & $0.1566$\\
		                           &OCS &  $3.2712\times10^{-8}$ & $0.0945$ &                            & $1.9786\times10^{-18}$ & $0.1041$\\
		                           &PSO & $1.7311\times10^{-28}$ & $0.1407$ &                            & $6.7831\times10^{-17}$ & $0.1238$\\
		\hline
		\multirow{3}{*}{$f_{22}$}  &OCD &                    $0$ & $0.0139$ &  \multirow{3}{*}{$f_{27}$} &  $6.6267\times10^{-5}$ & $0.0866$\\
		                           &OCS &                    $0$ & $0.0236$ &                            &  $4.3602\times10^{-7}$ & $0.0666$\\
		                           &PSO & $1.9093\times10^{-20}$ & $0.0815$ &                            &  $4.4452\times10^{-8}$ & $0.1052$\\
		\hline
		\multirow{3}{*}{$f_{23}$}  &OCD &                    $0$ & $0.0638$ &  \multirow{3}{*}{$f_{28}$} &                    $0$ & $0.0271$\\
		                           &OCS &                    $0$ & $0.0587$ &                            &                    $0$ & $0.0340$\\
		                           &PSO &                    $0$ & $0.0956$ &                            &               $5.0524$ & $0.0800$\\
		\hline
		\multirow{3}{*}{$f_{24}$}  &OCD &            $2444.2318$ & $0.9708$ &  \multirow{3}{*}{$f_{29}$} &                    $0$ & $0.0362$\\
		                           &OCS &            $2521.9434$ & $0.6238$ &                            &                    $0$ & $0.0442$\\
		                           &PSO & $6.4308\times10^{-24}$ & $0.5506$ &                            &                    $0$ & $0.0894$\\
		\hline
		\multirow{3}{*}{$f_{25}$}  &OCD & $2.0377\times10^{-11}$ & $0.0222$ &  \multirow{3}{*}{$f_{30}$} & $1.7764\times10^{-15}$ & $0.0526$\\
		                           &OCS & $2.0377\times10^{-11}$ & $0.0348$ &                            &                    $0$ & $0.0441$\\
		                           &PSO & $2.0377\times10^{-11}$ & $0.0882$ &                            &                    $0$ & $0.0993$\\
		\hline
	\end{tabular}
\end{table}

\subsection{30-dimensional case}\label{Sec4.3}

In the $30$-dimensional case, only the results of the OCS and PSO schemes are compared, since the OCD scheme is not applicable. For the OCS scheme, the number of iterations is set to $2000$ and the parameter $\lambda$ is set to $0.98$. The number of samples per iteration is $N=1000$, which corresponds to a total of $2000\times 1000=2000000$ evaluations of each objective function. For the PSO scheme, the parameters $w=0.5$ and $c_1=c_2=1.5$. The number of iterations is set to $500$ and the number of particles is set to $2000$ which correspond to a total of $500\times 2000=1000000$ evaluations of each objective function.  

\begin{figure}[htbp]
	\centering
	\includegraphics[width=\linewidth]{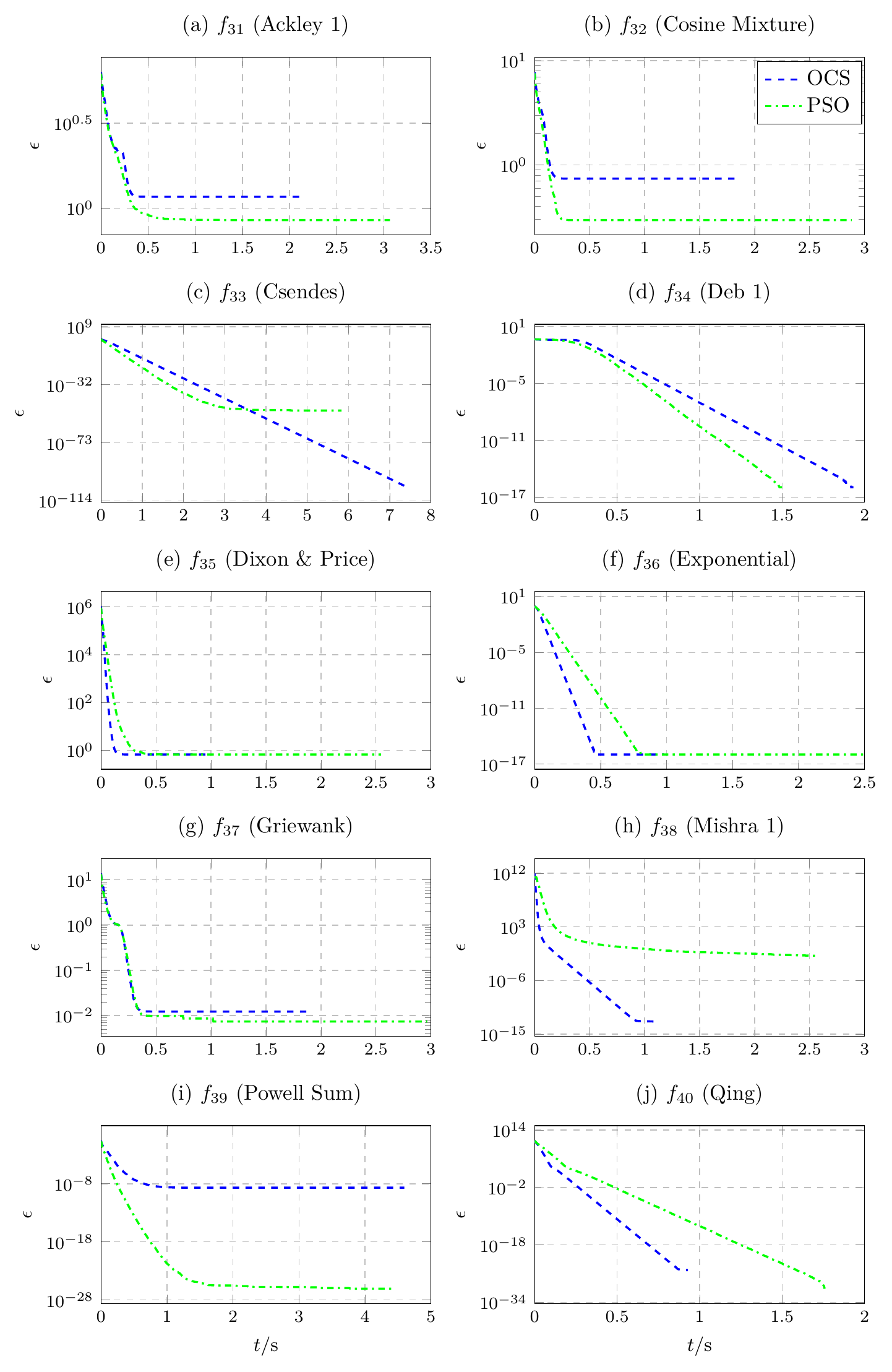}
	\caption{n-dimensional functions}
	\label{Fig_nd_1}
\end{figure}

\begin{figure}[htbp]
	\centering
	\includegraphics[width=\linewidth]{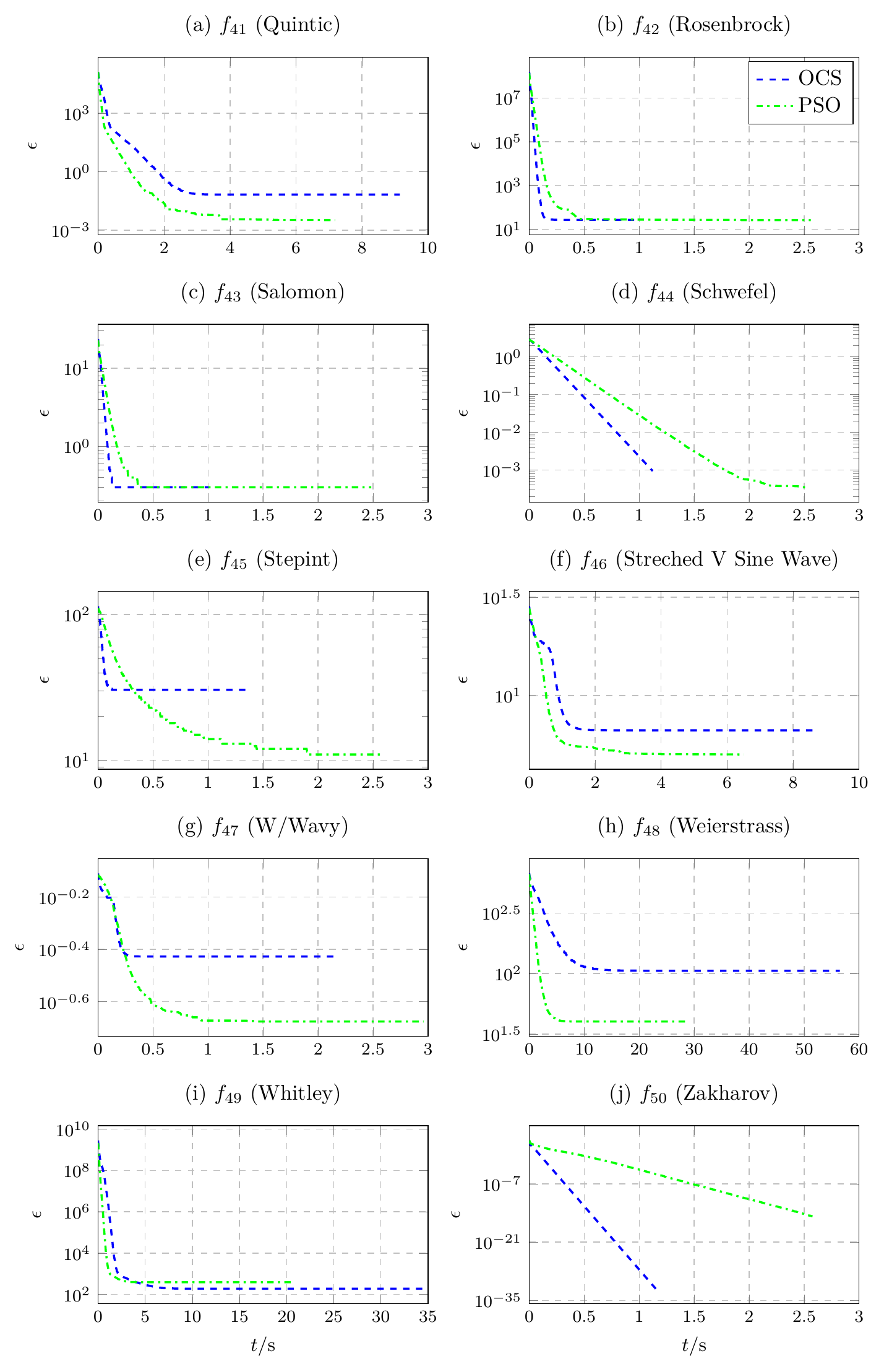}
	\caption{n-dimensional functions}
	\label{Fig_nd_2}
\end{figure}

The evolution of the error $\epsilon$ with respect to the computational time $t$ is presented in Fig. \ref{Fig_nd_1} and \ref{Fig_nd_2}. The final error and the computational time are collected in Tab. \ref{Tab_nd} for ease of comparison. From Fig. \ref{Fig_nd_1} and \ref{Fig_nd_2}, we see that in this case, both the OCS and PSO schemes can be easily trapped into a local minimum, especially when the objective function has lots of local minima. Moreover, compared to the PSO, it seems that the proposed OCS scheme tends to be trapped into a worse local minimum. From our point of view, the reason is that a swarm of particles is more flexibility than a single rectangular region. However, if a global minimum can be located, then the proposed OCS still converge faster than the PSO in most cases. In addition, it can be seen from Tab. \ref{Tab_nd} that the OCS is more efficient for functions that are easy to evaluate (which correspond to functions with lower computational time in the table). This is in sharp contrast to the Bayesian optimization which is more suitable for functions that are hard to evaluate. 

\begin{table}[htbp]
	\caption{Final error and computational time ($30$-dimensional cases)}
	\label{Tab_nd}
	\centering
	\begin{tabular}{llll|lll}
		\hline
		                          &    &             $\epsilon$ &    $t/s$ &                           &             $\epsilon$ &     $t/s$\\
		\hline
		\multirow{2}{*}{$f_{31}$} &OCS &               $1.1672$ & $2.1250$ & \multirow{2}{*}{$f_{41}$} &               $0.0670$ &  $9.1476$\\
		                          &PSO &               $0.8509$ & $3.0647$ &                           &               $0.0033$ &  $7.1850$\\
		\hline
		\multirow{2}{*}{$f_{32}$} &OCS &               $0.7389$ & $1.8674$ & \multirow{2}{*}{$f_{42}$} &              $26.7760$ &  $0.9567$\\
		                          &PSO &               $0.2956$ & $2.8825$ &                           &              $26.0295$ &  $2.5608$\\
		\hline
		\multirow{2}{*}{$f_{33}$} &OCS &$4.0252\times10^{-105}$ & $7.4280$ & \multirow{2}{*}{$f_{43}$} &               $0.2999$ &  $1.0038$\\
		                          &PSO & $6.4725\times10^{-51}$ & $5.8337$ &                           &               $0.2999$ &  $2.4848$\\
		\hline
		\multirow{2}{*}{$f_{34}$} &OCS &                    $0$ & $3.7868$ & \multirow{2}{*}{$f_{44}$} &  $9.4634\times10^{-4}$ &  $1.1233$\\
		                          &PSO &                    $0$ & $3.9395$ &                           &  $3.4933\times10^{-4}$ &  $2.5330$\\
		\hline
		\multirow{2}{*}{$f_{35}$} &OCS &               $0.6667$ & $0.9560$ & \multirow{2}{*}{$f_{45}$} &              $30.5000$ &  $1.3904$\\
		                          &PSO &               $0.6672$ & $2.5456$ &                           &                   $11$ &  $2.5996$\\
		\hline
		\multirow{2}{*}{$f_{36}$} &OCS & $1.1102\times10^{-16}$ & $0.9779$ & \multirow{2}{*}{$f_{46}$} &               $6.6715$ &  $8.6658$\\
		                          &PSO & $1.1102\times10^{-16}$ & $2.5115$ &                           &               $5.0310$ &  $6.5056$\\
		\hline
		\multirow{2}{*}{$f_{37}$} &OCS &               $0.0123$ & $1.8930$ & \multirow{2}{*}{$f_{47}$} &               $0.3740$ &  $2.1925$\\
		                          &PSO &               $0.0074$ & $2.9656$ &                           &               $0.2107$ &  $2.9587$\\
		\hline
		\multirow{2}{*}{$f_{38}$} &OCS & $1.2390\times10^{-13}$ & $1.1133$ & \multirow{2}{*}{$f_{48}$} &             $105.5264$ & $56.5051$\\
		                          &PSO &               $0.0137$ & $2.5571$ &                           &              $40.1300$ & $28.4343$\\
		\hline
		\multirow{2}{*}{$f_{39}$} &OCS & $2.1256\times10^{-09}$ & $4.5957$ & \multirow{2}{*}{$f_{49}$} &             $190.5336$ & $35.4969$\\
		                          &PSO & $8.9848\times10^{-27}$ & $4.4360$ &                           &             $392.3549$ & $20.4798$\\
		\hline
		\multirow{2}{*}{$f_{40}$} &OCS & $8.9634\times10^{-26}$ & $0.9288$ & \multirow{2}{*}{$f_{50}$} & $6.3418\times10^{-33}$ &  $1.1509$\\
		                          &PSO &                    $0$ & $2.4371$ &                           & $1.5273\times10^{-15}$ &  $2.5765$\\
		\hline
	\end{tabular}
\end{table}

\section{Conclusions}\label{Sec5}

In this paper, a new frame work for global optimization is proposed. Although it has a very simple form, it is effective for most of the benchmark functions. Compared to the PSO, the proposed frame work can generate new sample points at a lower computational cost. However, when the objective function is of high dimension and has lots of local minima, it usually requires more evaluations to avoid getting trapped into a local minimum since the utilization efficiency of samples is also lower. The reason is that in generating new sample points, only the information of the current best point is used. In this regard, it is more suitable for functions that are easy to evaluate which is in sharp contrast to the Bayesian optimization. On the other hand, if a global minimum can be located, then the proposed frame work usually converge faster than the PSO. From this point of view, the proposed frame work is more suitable for functions with a moderate dimension or a moderate number of local minima.

Future works should be focus on improving the utilization efficiency of sample points and preventing the algorithm from being trapped into a local minimum. From our perspective, a promising direction is to maintain multiple rectangular subregions in each iteration.

\section*{References}

\bibliography{mybibfile}

\end{document}